\documentclass[preprint2]{aastex631}

\begin{document}

\title{Galaxy-Absorber Association in the Epoch of Reionization: Galactic Population Luminosity Distribution for Different Absorbers at $10\geq z \geq5.5$}

\author[0000-0002-0761-1985]{Samir Ku\v{s}mi\'{c}}
\affiliation{New Mexico State University \\
MSC 4500, PO BOX 30001 \\
Las Cruces, NM 88003}
\email{samirk@nmsu.edu}

\author[0000-0002-0496-1656]{Kristian Finlator}
\affiliation{New Mexico State University \\
MSC 4500, PO BOX 30001 \\
Las Cruces, NM 88003}
\affiliation{Cosmic Dawn Center (DAWN) \\
Niels Bohr Institute, University of Copenhagen / DTU-Space \\ 
Technical University of Denmark}

\author[0009-0004-8503-0483]{Ezra Huscher}
\affiliation{New Mexico State University \\
MSC 4500, PO BOX 30001 \\
Las Cruces, NM 88003}

\author[0009-0002-0435-5055]{Maya Steen}
\affiliation{New Mexico State University \\
MSC 4500, PO BOX 30001 \\
Las Cruces, NM 88003}



\begin{abstract}

%ABSTRACT HERE. \lipsum[1-1]

How do galaxies of different luminosities contribute to the metal absorber populations of varying species and strength? We present our analysis of the predicted metal contributions from galaxies as observed in quasar absorption line spectra during the end of the Epoch of Reionization (EoR; $10 \geq z \geq 5.5$). This was done by implementing on-the-fly particle tracking into the latest \textsc{Technicolor Dawn} simulation and then linking CII, CIV, SiII, SiIV, OI, and MgII absorbers to host galaxies in post-processing. We define the Host Galaxy Luminosity Distribution (HGLD) as the rest-frame ultraviolet luminosity distribution of galaxies contributing ions to an absorber, weighted by the fractional contribution, and compute its dependence on ion and absorber strength. The HGLD shape is predicted to be indistinguishable from the field luminosity function, indicating that there is no relationship between the absorber strength or ion and the luminosity of the dominant contributing galaxy. Switching from galaxy luminosity to stellar mass, the predicted host galaxy mass distributions (HGMD) indicate that more-massive galaxies contribute a higher fraction of metal ions to absorbers of each species, with the HGMD of stronger absorbers extending out to higher masses. We conclude that the fraction of absorbing metal ions contributed by galaxies increases weakly with stellar mass, but the scatter in luminosity at fixed stellar mass obscures this relationship. For the same reason, we predict that observational analyses of the absorber-galaxy relationship will uncover stronger trends with stellar mass than with luminosity.

\end{abstract}

%{Classical Novae (251) --- Ultraviolet astronomy(1736) --- History of astronomy(1868) --- Interdisciplinary astronomy(804)}


\section{Introduction} \label{sec:intro}

The processes of star formation and feedback remain pressing uncertainties in our understanding of metal enrichment as they dictate the baryon distributions and metal abundances in and near regions where gravitational instability takes over. A key probe of these processes is the abundance and spatial distribution of metals, which are the byproduct of massive star formation. 

For example, the volume-averaged density of metals constrains the cosmic star formation history~\citep[][]{Madau:2014}{}{}. A complete census of cosmic metals that avoids the limitations of emission-based galaxy surveys must include metals in the circumgalactic and intergalactic media (CGM and IGM, respectively;~\citealt{Peroux:2020}). To this end, a number of ground-based surveys have traced the evolving abundances of ions including MgII, SiIV, CIV, and OI into the Epoch of Reionization~\citep[EoR;][]{Davies:2023a,Davies:2023b, Galbiati:2023,D'Odorico:2022, Cooksey:2013, Chen:2010,Becker:2019}. Analogous efforts continue into the \emph{JWST} epoch~\citep{Bordoloi:2023,Christensen:2023}.

The extent to which observations of metals constrain star formation and feedback depends on the measurements that are available. Ensemble statistics such as the metal absorber column density distribution (CDD) and its integral, the metal density, constrain the cosmic stellar mass density but yield limited insight into star formation and feedback because the metals' source galaxies are generally unidentified~\citep[for example,][]{Hasan:2022}. These processes are more directly constrained by the physical association between galaxies and metal absorbers. Drawing such connections, in turn, requires theoretical modeling~\citep[for example,][]{Steidel:2010,Bouche:2012, Churchill:2013}.
%In recent years, \kfc{parallel survey efforts have identified} metal-line absorbers and extended the census of galaxies \sout{of lower masses} \kfc{into} the Epoch of Reionization (EoR). These surveys continue into the \emph{JWST} era \citep{Bordoloi:2023, Christensen:2023} \kfc{Add one or two JWST galaxy survey papers?} having been previously studied based on species and tracing their populations, including confirmed and extensively measured MgII, SiIV, and CIV \citep[][]{Davies:2023a,Davies:2023b, Galbiati:2023,D'Odorico:2022, Cooksey:2013, Chen:2010}. With the plethora of absorber data available we understand how they are linked to the galaxies that they must come from.

Cosmological simulations have achieved limited success both in reproducing observations and in exposing the ways in which the details of star formation and feedback regulate ensemble statistics including the equivalent width and column density distributions~\citep[][]{Oppenheimer:2008, Oppenheimer:2009, Finlator:2020, Keating:2016, Turner:2016}{}{}. In detail, however, models frequently overproduce weak CIV and SiIV absorbers and underproduce strong CIV systems at $z\geq4$~\citep{Rahmati:2016,Finlator:2020,Keating:2016,Hasan:2020}. The underproduction of strong CIV absorbers may reflect problems with the ionizing ultraviolet background (UVB;~\citealt{Keating:2016,Finlator:2020}), but the overproduction of weak systems has two possible interpretations. On the one hand, if absorber strength increases with host galaxy mass, then the surfeit of weak systems suggests that simulated low-mass galaxies expel too many metals. In support of this idea, simulations also predict that most CIV and OI absorbers are hosted by galaxies with absolute ultraviolet magnitudes $M_\mathrm{UV}$ fainter than -19~\citep{Finlator:2013,Finlator:2020,Doughty:2023}, and the recent detection of a bright galaxy associated with a strong OI absorber may conflict with this prediction~\citep{YunjingWu:2021}. On the other hand, it is possible that simulated massive galaxies simply eject their metals too far, endowing them with an unrealistically large geometric cross section to weak absorption. Distinguishing between these interpretations requires an improved understanding of the galaxy-absorber relationship. 

%The question of which galaxies generated the metals that are observed in absorption overlaps significantly with the question of which galaxies produced the bulk of the ionizing flux that drove HI reionization because the same stars that dominate cosmic metal production also dominate its ionizing emissivity. Indeed, if one assumes that the fractions of new metals and ionizing flux that galaxies emit into the IGM have the same mass dependence, then the questions are the same, emphasizing the importance of metal absorption to studies of reionization. 

%Efforts to identify observationally the relative roles of faint and bright galaxies remain controversial. The steep faint-end slope of the ultraviolet luminosity function (UVLF) suggests that, during the EoR, most stars formed in faint galaxies~\citep[][]{Bouwens:2009}. Cross-correlation of the IGM with CIV absorbers has likewise attributed reionization to an unobserved, faint population~\citep{Meyer:2019}. Efforts to identify the hosts of OI absorbers  at millimeter/submillimeter wavelengths have yielded upper limits in five out of six cases~\citep{Wu:2023}, with inconclusive implications for the possible role of faint galaxies. By contrast, other analyses have argued that moderately-massive~\citep[][]{Atek:2023a, Naidu:2020} or even massive galaxies~\citep[][]{Labbe:2023} may dominate star formation at early times.

Observational analyses usually assume that metal absorbers are physically-associated with the nearest observable galaxy (\citealt{Hasan:2022, Churchill:2013, Steidel:2010}; see, however~\citealt{Wu:2023}). While a useful assumption, this neglects the possible role of even fainter galaxies, whose presence is expected even if for no other reason than owing to galaxy clustering~\citep{Finlator:2020, Finlator:2013, Rahmati:2014}. Analyses of numerical simulations generally take a similar approach~\citep{Keating:2016,Finlator:2020} because it is difficult to infer from a simulation snapshot where metals that have long-since been ejected into the CGM were originally formed. In order to use observations of metals to constrain models of star formation and feedback, it is important to know not only where the metals are, but where they formed.

%Whereas efforts to detect galaxies in emission are inherently biased towards bright galaxies, the environments of less-massive galaxies are capable of being detected through absorption line spectroscopy. Detection of metals in absorption line spectroscopy is due a metal absorber cloud residing in the circumgalactic medium (CGM) of a galaxy, with detections confirmed out to $z=6$ \citep{Becker:2019, D'Odorico:2022, Davies:2023a, Christensen:2023}. Theoretical works show high spatial correlation between these absorbers and galaxies \citep{Finlator:2020, Doughty:2023} that coincides with observational efforts \citep[][]{Steidel:2010, Adelberger:2005}{}{}. As such, we can confirm that a galaxy is spatially near an absorption line, complementing the idea that absorbers and galaxies are in a linked system.

%This is the farthest extent of our knowledge with absorbers and it is impossible to observationally determine one's cloud source with direct detection. The paradigm may be intuitive in that the more-massive galaxies will hold a larger gas reservoir and more stars, so more feedback activities are possible, but it can be argued that the lower gravitational potential in less-massive and possibly unobservable galaxies allows more gas to escape into their CGM and challenges the paradigm. The host galaxy is ambiguous for absorbers within simulated protocluster/group environments. However, lowest-mass galaxies are not believed to be dominant metal sources within these systems \citep[][]{Kirby:2011}{}{}. We do not know if the paradigm is correct given our models.

In order to address this need, we have developed a particle-tracking analysis that quantifies the mass and luminosity distribution of galaxies that contribute metals to absorbers. In Section~\ref{sec:method}, we describe the simulation and our method for linking absorbers with galaxies. In Section~\ref{sec:results}, we present results of the analysis. In Section~\ref{sec:discuss} we interpret the results to our findings on the trend of galaxy-abosrber linkings, including whether stronger absorbers are closely linked to more luminous and more massive galaxies. Throughout this work, we assume flat $\Lambda \mathrm{CDM}$ with $ h = 0.6774 $, $\Omega_{\mathrm{m}} = 0.3089$, $ \Omega_{\mathrm{b}} = 0.0486$, and $ \Omega_{\Lambda} = 0.6911$. Readers who are interested in our analysis methods are referred to Section \ref{sec:method}. Readers interested in the host galaxy luminosity distribution (HGLD), the number of galaxies per luminosity bin per absorber, are referred to the definition in Section \ref{ssec:idea} and the results in Section \ref{sec:results}.
%\kfc{Include a figure of the "contributing galaxy" fraction and the escaped metal fraction, and reference the Kirby+2011 study?} \skc{Are you referencing the plots you made? I never got the PDF versions}\kfc{This should go in the discussion if anywhere; the intro is now quite nice as it is.}
\section{Analysis and Methodology} \label{sec:method}

\subsection{\textsc{Technicolor Dawn} Simulation} \label{subsec:sim}

Our radiation hydrodynamic simulation was run using the \textsc{Technicolor Dawn} \citep{Finlator:2018} code. It models the evolution of a $(12h^{-1})^3\:\mathrm{cMpc^3}$ volume using $(2 \times 512)^3$ particles. Gas particles initiate with masses of $2.563 \times 10^5\:\mathrm{M_{\odot}}$ and dark matter particles initiate with $1.372 \times 10^6\:\mathrm{M_{\odot}}$. The ionizing emissivity from star-forming gas is modeled using BPASS version 2.2.1 \citep{Stanway:2018}. This improvement with respect to previous work accounts for binary stellar evolution. It necessitates small changes to the ionizing escape fraction and initial mass function; see \citet[][]{Huscher:2024}{}{} for details. Ionizing flux is propagated into the simulation using a moment method in which the radiation field is discretized spatially into a uniform grid of $80^3$ voxels and spectrally into 24 frequency bins uniformly spanning 1--10 Ryd accounting for ionization into species observed thus far. The contribution from QSOs is accounted for using a volume-averaged calculation; see~\citet{Finlator:2020} for details. Star-forming gas is ejected into a galactic wind via a Monte Carlo model in which the velocity and mass-loading factor are anchored in high-resolution simulations; see~\citep{Finlator:2020} for details. Following ejection, gas and metals circulate in the circumgalactic medium, where they compose the bulk of observable metal absorbers.

We identify galaxies with spline kernel interpolative denmax \citep[\textsc{SKID,}][]{Governato:1997}{}{} in star-gas mode. Synthetic photometry is generated as follows: we obtain each star particle's stellar continuum by interpolating to its metallicity and age within the BPASS Version 2.2.1 models \citep{Stanway:2018} using binary stars with the \textsc{imf\_135\_100} initial mass function and summing all stellar continua within each galaxy. Dust is ignored. We then weight the stellar continuum by an idealized, rounded tophat filter running from 1300--1700\AA~to compute the UV continuum luminosity at a central wavelength of 1510\AA.

\begin{figure*}[t]
    \centering
    \includegraphics[width=\textwidth]{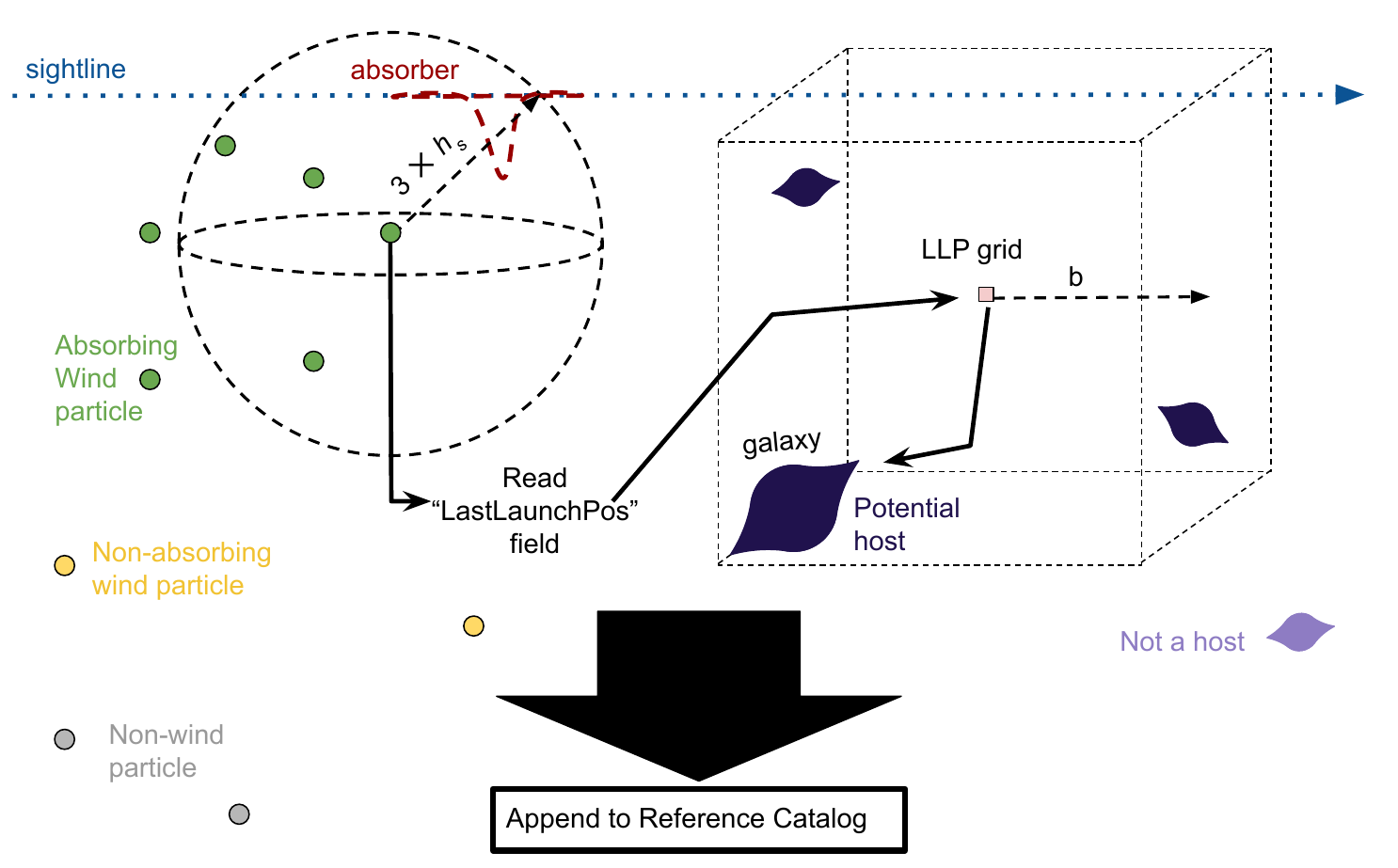}
    \caption{ Diagram illustrating our ``\textsc{hostgal}" analysis. First we map the absorber's position (red, dashed bump) from velocity space along the sightline (blue, dotted arrow) into physical coordinates. Then, we identify gas particles (green dots) that lie close enough to contribute based on their smoothing lengths $h_s$. If a gas particle is flagged as a wind particle, then its ``LastLaunchPos" (LLP) is extracted and we search the LLP grid cell looking around a cubical search buffer $b$. We link galaxies lying within that buffer to the gas particle and absorber and appended to an output reference catalog.}
    \label{fig:visual_hostgal}
\end{figure*}

We extract an absorber catalog using a quasar sightline caster following \citet[][]{Finlator:2020}. The oblique quasar sightline periodically wraps the simulation box until it spans a given velocity distance. We discretize the sightline into pixels and compute the local density, temperature, proper motion, metallicity, and ionizing background using the local SPH particles. Our use of the local UVB enables us to account realistically for small-scale UVB fluctuations. We use the assumption of ionization equilibrium to compute all metal ionization fractions at each pixel. We project from physical into observed space, using Voigt profiles to compute the optical depth at each velocity bin and account for simulated proper motions. We finalize the spectrum by convolving the intrinsic (that is, ``ideal") spectrum with a Gaussian function to emulate instrument response and adding noise. We then march a 3-pixel window along the resulting spectrum to find $5 \sigma$ drops of transmission and flag them as absorption features, emulating observational efforts.

%\textcolor{green}{[HERE IS HOW I WOULD WRITE THAT LAST PARAGRAPH. TAKE IT IN WHOLE OR IN PART.] We extract a catalog of synthetic absorbers from our simulations following the technique described in~\citet{Finlator:2020}. Briefly, we cast a sightline through the simulation that is oblique to its boundaries, wrapping periodically around it until it spans a given velocity distance. We discretize the sightline into pixels and compute the local density, temperature, proper motion, metallicity, and ionizing background using the local SPH particles. Our use of the local UVB enables us to account realistically for small-scale UVB fluctuations. We use the assumption of ionization equilibrium to compute all metal ionization fractions at each pixel. We project from physical into observed space, using Voigt profiles to compute the optical depth at each velocity and accounting for simulated proper motions. We filter the result using a realistic instrument response function and add Gaussian noise. Finally, we extract absorbers from the synthetic quasar sightline in a way that mimics observational efforts.}

\subsection{The Host Galaxy Luminosity Distribution}\label{ssec:idea}

We wish to compute the HGLD, which is the fractional mass contribution $\frac{df}{dM}(M|N)$ from galaxies with luminosity $M$ to metal absorbers of a particular column density $N$. For a hypothetical absorber consisting of a single gas particle whose host galaxy $i$ is known precisely, this is simply a $\delta$-function $\frac{df}{dM} (M) = \delta(M_i-M) = 1$ if $M_i-M = 0$. In practice, our particle-tracking approach does not uniquely match each wind particle to a host galaxy, but rather to the set of $N_\mathrm{host}$ galaxies lying close to the position from which it was last launched. Consequently, a single-particle absorber's HGLD becomes an average over multiple $\delta$-functions:
\begin{equation}
\frac{df}{dM}(M) = \frac{1}{N_\mathrm{host}} \sum_{i=1,N_\mathrm{host}} \delta(M_i-M)
\label{eqn:1PartManyHosts}
\end{equation}
where $i$ runs over all potential host galaxies for that particle.

Realistic absorbers consist of more than one gas particle, and we average over many such absorbers within a narrow range of column densities in order to improve our statistics. Accounting for these adjustments, the mean HGLD becomes

\begin{equation}
\frac{df}{dM}(M|N_{\mathrm{col}}) \equiv \frac{1}{N_\mathrm{part}}\sum_{j=1,N_\mathrm{part}} \frac{1}{N_\mathrm{host,j}} \sum_{i=1,N_{\mathrm{host},j}}\delta(M_{i,j}-M)
\label{eqn:hgld_norm}
\end{equation}

where $j$ runs over all $N_\mathrm{part}$ gas particles contributing to any absorber with column density $N_{\mathrm{col}}$; $i$ runs over all $N_{\mathrm{host},j}$ candidate host galaxies (hereafter, ``hosts") for the $j$th particle; and $M_{i,j}$ is the luminosity of the $i$th host for the $j$th gas particle. The mean number of hosts per absorber per luminosity at a given column density results from a slightly different normalization:

\begin{equation}
\frac{dn}{dM}(M|N_{\mathrm{col}}) \equiv \frac{1}{N_\mathrm{abs}(N_{\mathrm{col}})}\sum_{j=1,N_\mathrm{part}} \sum_{i=1,N_\mathrm{host},j}\delta(M_{i,j}-M)
\label{eqn:hgld_unnorm}
\end{equation}

Summing $\frac{dn}{dM}$ over all $M$ reveals the mean number of hosts per absorber with column density $N_{\mathrm{col}}$. Analogous relations may be used to study the distribution of host galaxy in mass rather than luminosity. In Section~\ref{ssec:method}, we describe how each gas particle's hosts are identified.

\subsection{Particle Tracking and Galaxy-Absorber Linking}\label{ssec:method}

\begin{figure}
    \centering
    \includegraphics[width=0.45\textwidth]{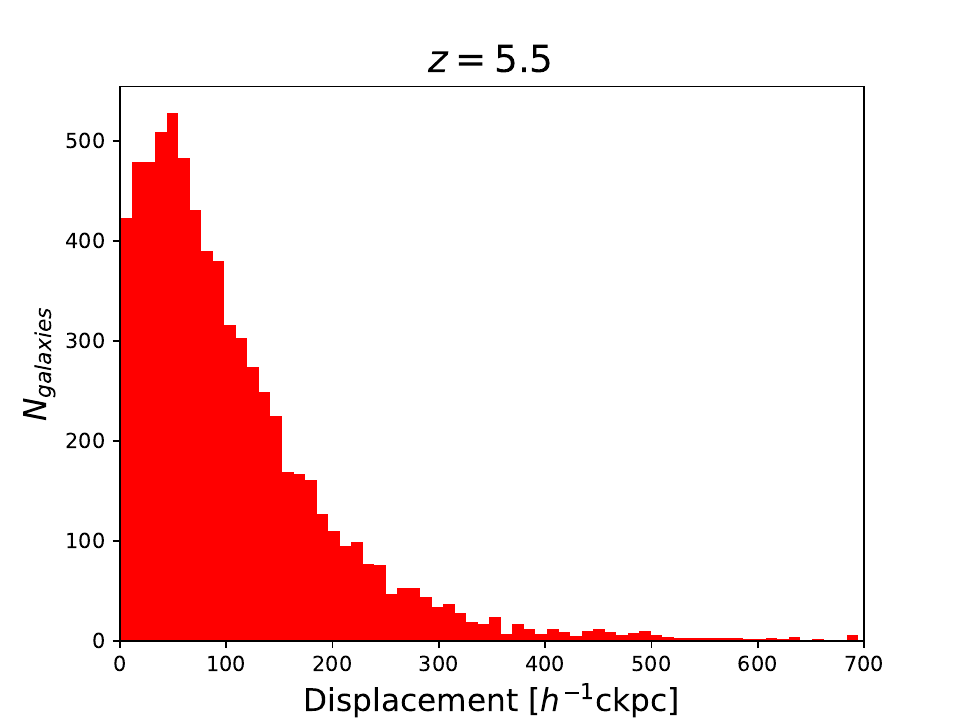}
    \caption{Histogram showing the displacement between each ``LastLaunchPos" grid cell from its center to the nearest galaxy.}
    \label{fig:gal_disp} 
\end{figure}

\begin{figure*}
    \centering
    \includegraphics[width=1.0\textwidth]{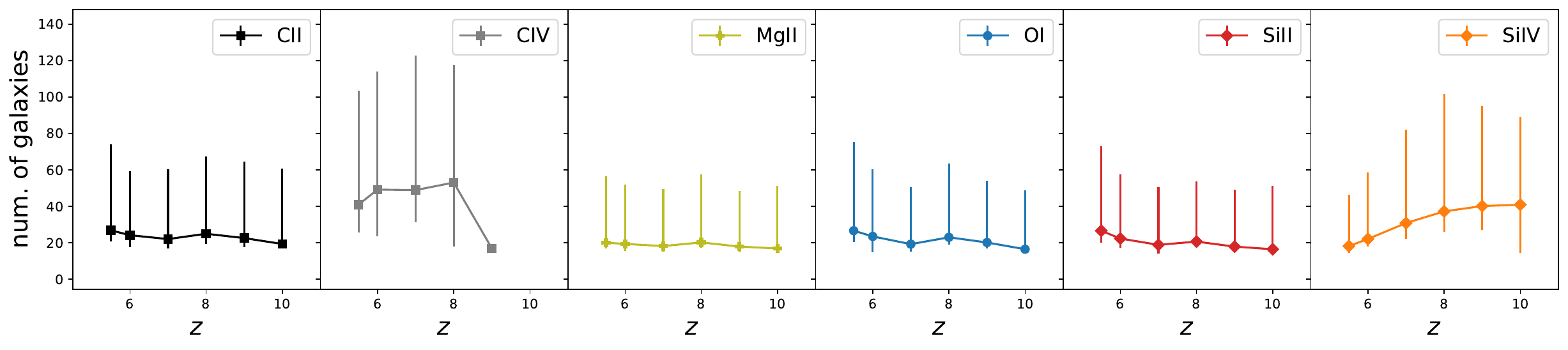}
    \caption{Number of host galaxies linked per gas particle by \textsc{hostgal} at several redshifts. The point is the mean value and error bars extend out to 1$\sigma$.}
    \label{fig:gals_per_gaspart}
\end{figure*}

Our goal is to relate extragalactic metals to the host galaxies that launched them in post-processing. To this end, two new fields were added to each baryon particle, namely ``LastLaunchPos" (LLP) and ``LastLaunchTime'' (LLT), which respectively store the position and scale factor of the last launch instance the particle experienced. ``LastLaunchPos" is saved as an integer encoding the indices for the grid space of the simulation split into $1625^3$ cells to conserve memory resources. If this particle is re-accreted and re-launched, then both fields are overwritten. 

To link these galaxies and absorbers in post-processing, a new technique was developed into a software package called \textsc{hostgal} under \textsc{pycosie} \citep[][]{kusmic_2024_pycosie}, our host galaxy finder for gas particles. For each snapshot, we identify gas particles that have been flagged as launched, hereafter called ``wind particles." We verify that each wind particle has not been re-accreted into a galaxy's star-forming medium and that it is enriched. Next, we determine whether it contributes to any absorber of various species within our simulated quasar sightline. We use each particle's unique smoothing length, which is typically 37--41 ckpc/h from $z=10$ to $5.5$, to characterize how it fills its local volume using its SPH smoothing kernel. If a wind particle lies within four times its smoothing length to an absorber and contains metals matching the species of the absorber, then we determine the grid cell from which the particle was most recently launched. Owing to galaxy proper motions and to the coarse nature of our LLP grid, this information does not uniquely match each gas particle with a host galaxy. We identify all galaxies within a cubical region centered on the LLP as potential hosts . Since we do not have information other than when and where the gas particle has been launched, we include all galaxies within a search window of fixed comoving size as potential hosts. We verify that selecting the host galaxy as the closest galaxy to the LLP grid cell's center does not impact our results significantly as seen in Appendix \ref{app:search_box}. Additionally, our search window approach is more robust as it suppresses misidentification in exchange for some contamination by unassociated galaxies. The software outputs a reference table linking the absorber, species, gas particle, and host galaxies. Figure~\ref{fig:visual_hostgal} illustrates our approach graphically.

We compute a comoving search volume about each gas particle's LLP that is large enough to include the particle's unknown true host galaxy but small enough to minimize contamination. To account for the possibility that galaxies with large proper motions have moved significantly since launching a gas particle, we determined the closest galaxy to each LLP grid center, calculated the distances between the them, and derived the distribution of these distances as seen in Figure \ref{fig:gal_disp} to obtain a distribution of possible displacements. This allows us to include isolated field galaxies and calculate robustly the distance a galaxy can travel from the LLP irrespective of their local environment.
We find that 95\% of the distance distribution is contained within a distance of 294 ckpc/h at $z=5.5$ and adopt this as our search buffer (half of the search box's side length).
%\kfc{Do we really need to include this next qualification? If it's good enough, maybe we don't have to. If it's important, then we should probably go all-in and describe it in an Appendix.} Although the test run used previously to get the distribution is smaller than the run used for this analysis, the mass resolutions are similar and we do not expect large enough deviation in properties between each runs' galaxy populations to be impactful. A visual of the distribution is provided in Figure \ref{fig:gal_disp}.

%\kfc{The next two paragraphs basically describe in prose what Equation 2 is doing because that implicitly includes linking incidences. Would it be enough just to describe this in prose? You might even be able to eliminate the last paragraph completely.} It feels important to define our weights: justification, how it's made, what it assumes
A drawback to our approach for deriving each absorber's HGLD is that many galaxies are usually included in our search region. This can be seen in Figure \ref{fig:gals_per_gaspart}, which shows there are routinely 10--100 galaxies located within the search region centered about each gas particle's LLP. To combat the ambiguity, we weight the galaxy-absorber links by what we will refer to as linking incidences, which is the number of times a unique galaxy is present within the reference table. This weights more towards galaxies that appear in the table more often with the assumption that galaxies that appear more have contributed more gas. These weights are applied to HGLD binning for each galaxy on a per-absorber basis. 

\begin{figure*}[t]
    \centering
    \includegraphics[width=\textwidth]{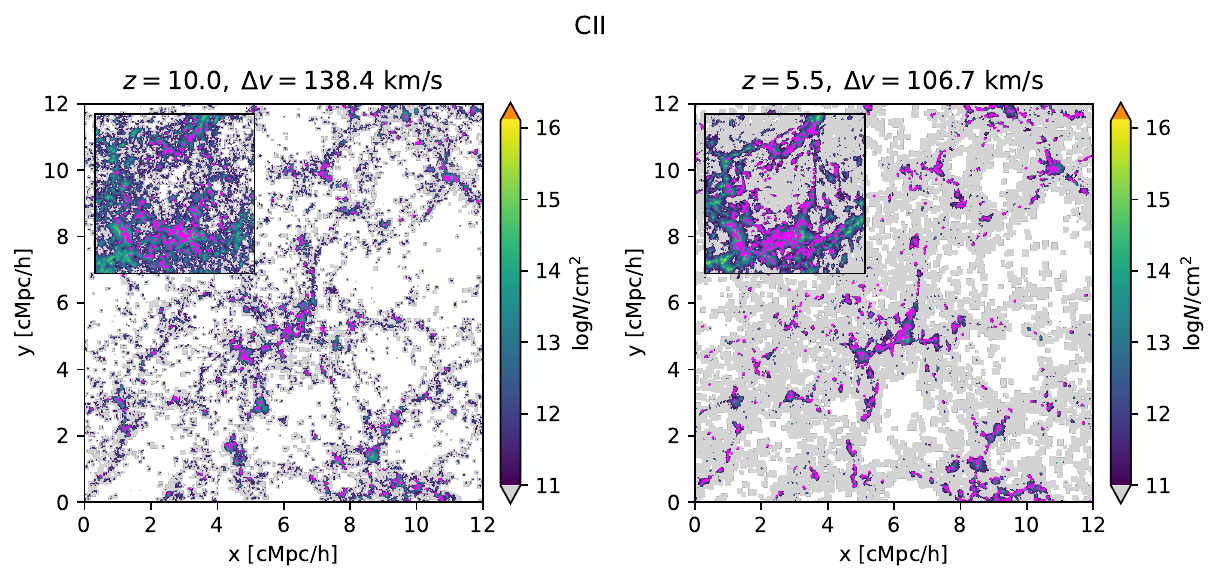} 
    \includegraphics[width=\textwidth]{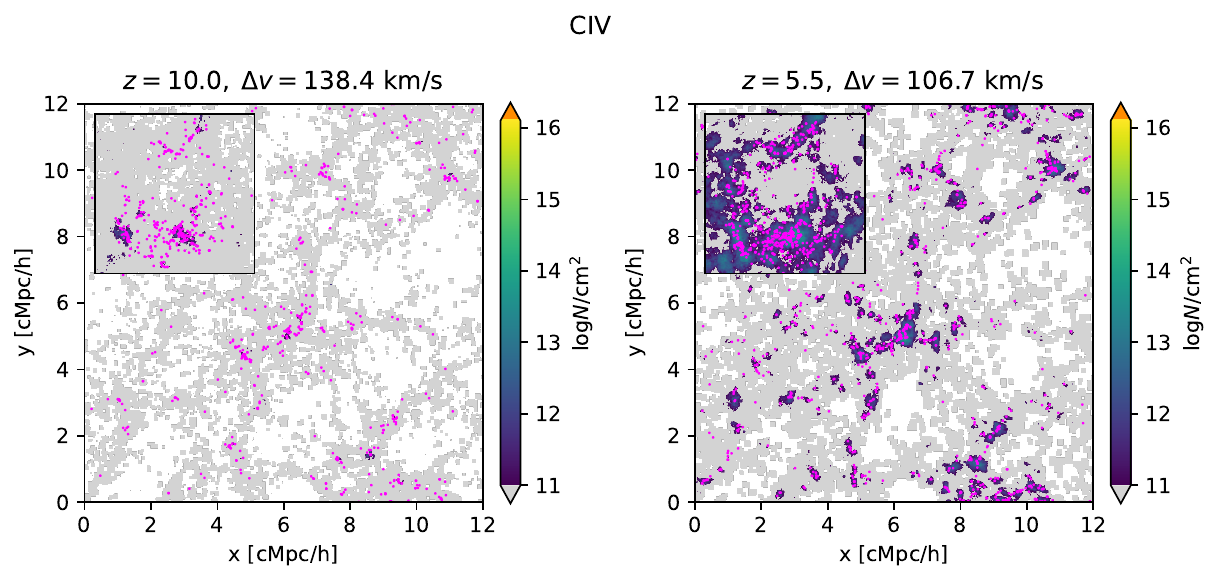}
    \caption{Slice of the simulation with calculated column densities for CII and CIV in a $512\times512\times16$ comoving grid for $x$, $y$, and $z$ respectively. Light gray areas indicate where the ion column density $\log N /\mathrm{cm}^{-2} < 11$ while orange areas are where $\log N /\mathrm{cm}^{-2} > 16.1$. The inset is a zoomed-in sphere of the center of the slice with radius of 2000 ckpc/h. Each plot is titled with the redshift of the snapshot and the depth of the slice in km/s. The magenta dots are galaxies.
    }
    \label{fig:CII_CIV_maps}
\end{figure*}

\begin{figure*}[ht]
    \centering
    \includegraphics[width=0.417\textwidth]{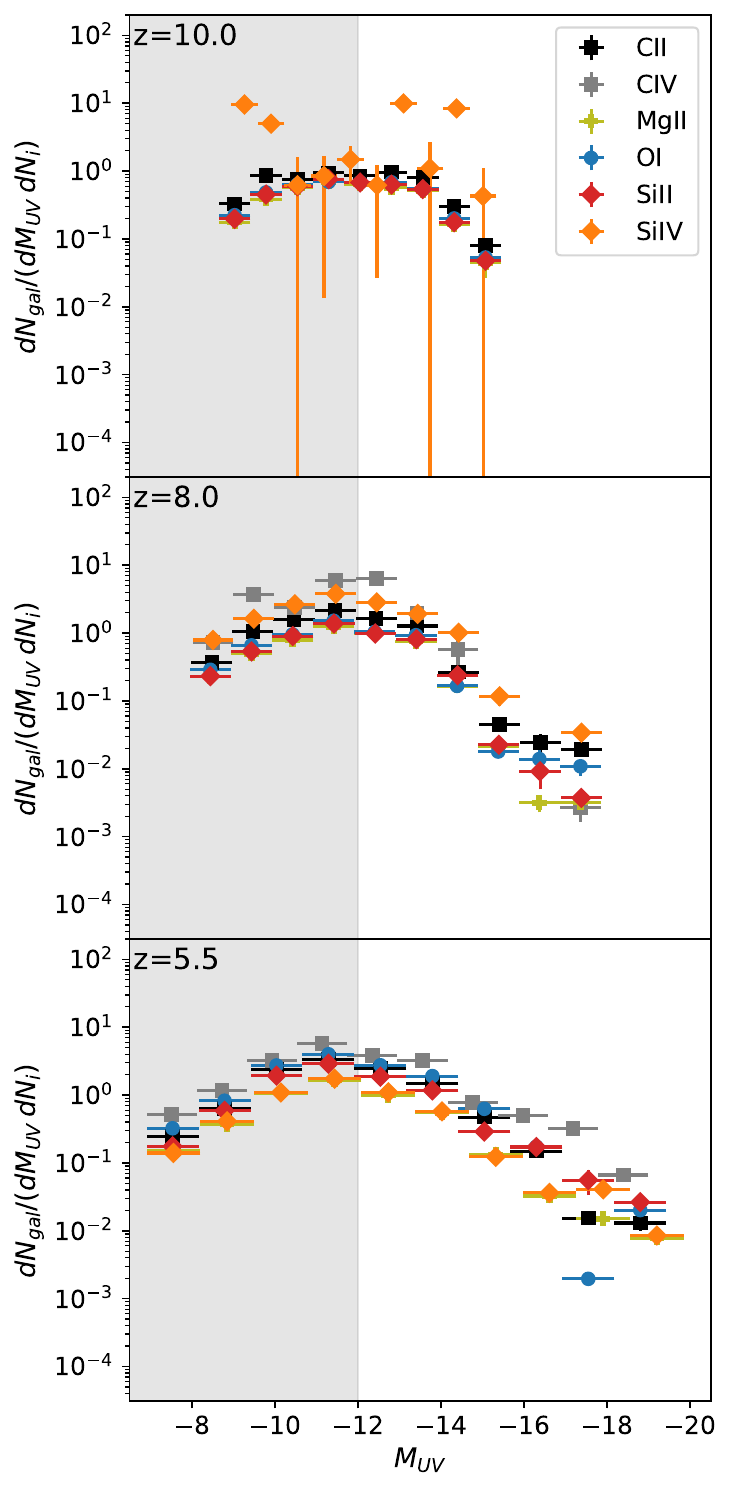} 
    \includegraphics[width=0.4\textwidth]{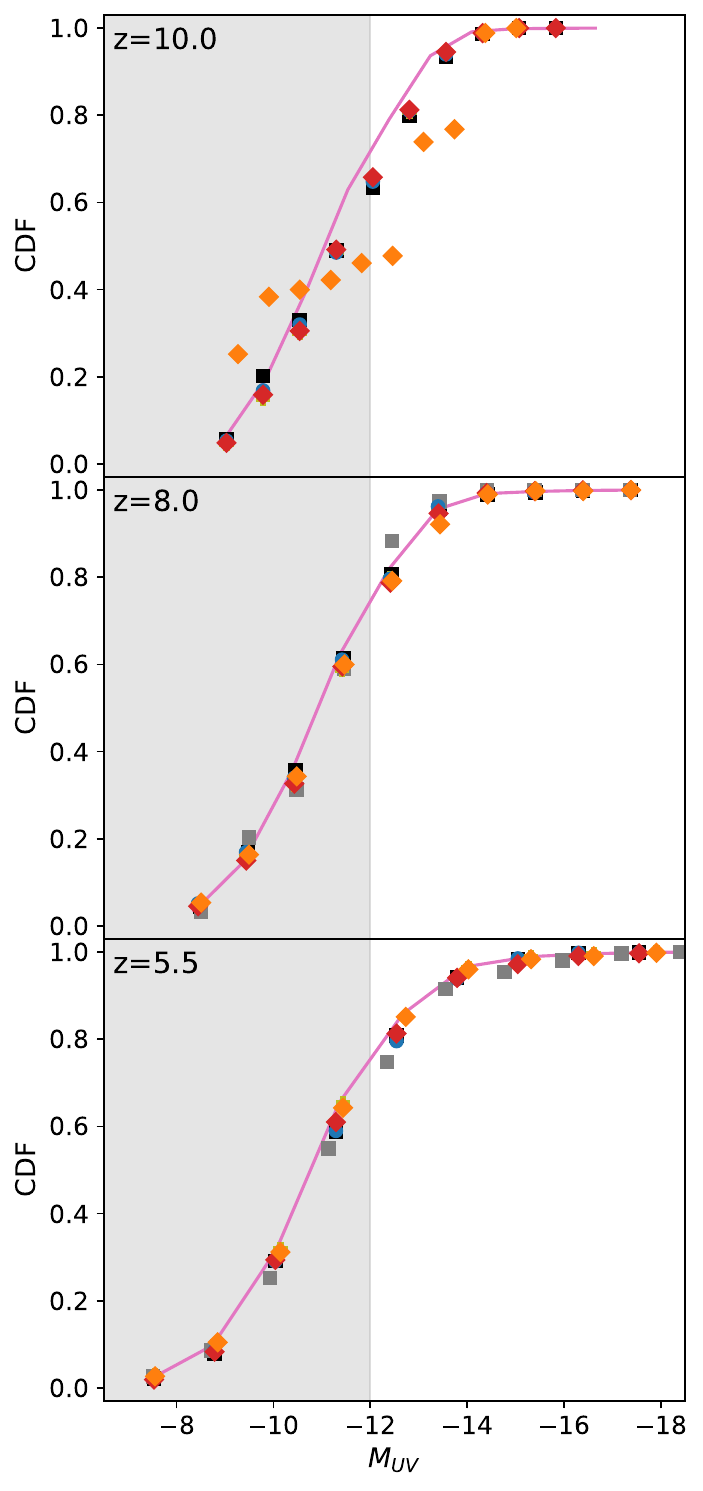}
    \caption{Host galaxy UV luminosity distribution throughout each redshift for each species (Equation~\ref{eqn:hgld_unnorm}; left) and their cumulative integrals compared to the cumulative field UVLF (right). The grey shaded region is the resolution limit of the simulation. This analysis does not indicate a luminosity-dependence for absorbers' host galaxies.}
    \label{fig:UVLF_all_W_pl12_CDF}
\end{figure*}

Our analysis begins with the distribution of host galaxies. As described in Section~\ref{ssec:idea}, the HGLD is normalized by the number of simulated absorbers of a given species as well as the corresponding linking incidences. This is achieved by dividing the reference table first by each species and finding each unique absorber. We then fetch the potential host galaxies and search them on the catalogs to find their $M_{\mathrm{UV}}$ and group them with each species to bin into histograms with bin widths determined using the Freedman-Dianconis rule.

\section{Results} \label{sec:results}

\subsection{Visual Correlation}

For a visual impression of the absorber-galaxy relationship, we show in Figure \ref{fig:CII_CIV_maps} the column densities of CII and CIV within a central slice of the simulation of thickness $\Delta v = (138.4, 106.7)$ km/s for $z = (10, 5.5)$ with galaxies as scatter points. At $z=10$ CII is already ubiquitous at column densities above $10^{11}$cm$^{-2}$, but CIV is just emerging at lower column densities owing to its low early ionization fraction~\citep{Finlator:2015}, itself a consequence of the weak early UVB. As time goes on, both species have increasing filling factors, but most of the growth is at column densities below $10^{11}$cm$^{-2}$. By $z=5.5$, CIV and CII occupy similar spaces with similar column densities.

Comparing the two-dimensional histograms with the galaxies (magenta points) indicates that CII and CIV consistenty correlate spatially with galaxies spatially at $z\sim10$ to 5.5. This reflects co-evolution between the ionization and metal filling factor: at $z=10$, the Universe remains mostly neutral, hence carbon is predominantly singly- (CII) or doubly-ionized (CIII) and is more prominent around galaxies. By $z=5.5$, ionizing flux becomes strongest near galaxies, which leads further ionizing to CIV around the present C regions to more prevalent column densities. This is consistent with correlations seen observationally \citep[][]{Wu:2023}{}{}; however, we did not further look into this and continued with our more direct method of galaxy-absorber association.

%\Motivating questions: What is the distribution of luminosities of galaxies that contribute metals to absorbers, and how does this distribution compare to the volume-averaged luminosity function?}

\subsection{Host Galaxy Luminosity Distribution}
%\kfc{Now that you're separately analyzing the HGLD and the HGMD, rearrange the presentation and section titles to reflect this. For example, The current Section 3.3 does address the ``Host Galaxy Distribution" and should be a subsection of the current Section 3.2, but if you create separate HGLD and HGMD sections, then current 3.3 could go into the HGMD section.} \label{sub:host_gal_dist}
Our central goal is to compute the predicted HGLD and its dependence on ion and absorber strength (Equation~\ref{eqn:hgld_unnorm}). We show the result in Figure~\ref{fig:UVLF_all_W_pl12_CDF}. The panels on the left-hand side indicate that each luminosity bin contributes equally to the absorber population. On the right-side panel of Figure \ref{fig:UVLF_all_W_pl12_CDF} is the cumulative distribution functions (CDF) of each species' HGLD compared to the field UV luminosity function (UVLF). Other than from low sample statistics, the shapes of the CDFs do not appear to differ from that of the field UVLF. This impression is confirmed by Kolmogorov-Smirnov (KS) tests as seen in Tables \ref{tab:KS_all_species_10} and \ref{tab:KS_all_species_5.5}, which compare each species to the field. These tests resulted in low coefficients with high $1-p$ confidence, aside from the low count statistics from CIV early on.

%\kfc{PARAGRAPH ABOUT HGLD NORMALIZATION}
Turning from the HGLD shape to its normalization (Equation~\ref{eqn:hgld_unnorm}), we find that its amplitude varies between ionic species. At $z=10$, the CII HGLD's amplitude is slightly higher that those of the other ions. The SiIV HGLD is noisy and the CIV HGLD is not computed because these ions are not abundant enough to yield statistically meaningful predictions. By $z=5.5$, the relative HGLD amplitudes rearrange so that they appear, in decreasing order, as CIV, OI, CII, SII and MgII, and then SIV. These amplitudes are consistent with previous correlation function amplitudes reported in \citet[][]{Doughty:2023}{}{}. 

One potential problem with our approach is that the number of host galaxies assigned to each absorber ($N_{\mathrm{host},j}$ in Equation~\ref{eqn:hgld_unnorm}) depends on the search region size, which in turn reflects a trade-off: a search region that is too large will include unassociated galaxies while a region that is too small may exclude the true host. We alleviate this with a sensible limit based on galaxy kinematics and the distance it can make with the LLP as described in Section~\ref{ssec:method}. However, the offset between the normalizations for each species should be insensitive to this issue. 

%\kfc{PARAGRAPH ABOUT HGLD NORMALIZATION}
The HGLD amplitude is higher for most stronger absorbers, suggesting that more galaxies are needed to produce a stronger absorber than a weaker absorber. This agrees with the cross-correlation amplitudes in \citet{Doughty:2023}, where integrating either informs how many galaxies are around each absorber, and both have boosted amplitudes for stronger absorbers. However, this is not true for highly-ionized species at $z=10$ to 8. 
%SiIV and CIV are unusual in that for a time period their stronger absorbers are more prominent among their absorber populations as seen in the bottom of Figure \ref{fig:UVLF_each_species}.\kfc{You haven't introduced this figure yet. Perhaps split this discussion into a separate paragraph and introduce F6 more deliberately?} High-ionization systems with weaker absorbers suggest richer environments as reionization progresses. The prevalence of stronger absorbers for high-ionization systems in these results suggest \textcolor{magenta}{more meager}\kfc{less overdense?} environments earlier on in cosmic time. If strong low-ionization absorber systems and weak high-ionization absorber systems mean a richer environment, then C and Si systems are around \textcolor{magenta}{more meager}\kfc{less overdense?} environments turning rich.\kfc{``meager" is not a standard technical word the way ``rich is."}

If the HGLD amplitude for a species is higher, then more galaxies are needed on average to make an absorber. This may be due to the difference in populations between the species that are present during this time~\citep{Davies:2023a, Finlator:2020}. Other wild variations are from low sample statistics with fewer absorbers present, such as with CIV at $z=10$. The turn-over at less luminous bins ($M_{\mathrm{UV}} \sim -12$) is due to the resolution limits from the simulation. However, the limits are fainter than \textit{JWST}'s limit of around (6.5, 7.5, 10) J-band magnitude for (high-resolution, medium resolution, PRISM) spectroscopy on the bright end \footnote{https://www.cosmos.esa.int/web/jwst-nirspec/brightness-limits}. Faint end results from Early Science releases show sensitivity that allow limiting absolute magnitudes of around -19 to -21 for $z=5$ to 10 \citep[][]{Harikane:2023}{}{}, so our UVLF turns over at a luminosity $10^9\times$ fainter even than JWST can go. This means that these abundant, unobservable galaxies are predicted to be the dominant contributors of reionization-epoch metal absorbers.

\begin{figure*}
    \centering
    \includegraphics[width=1.0\textwidth]{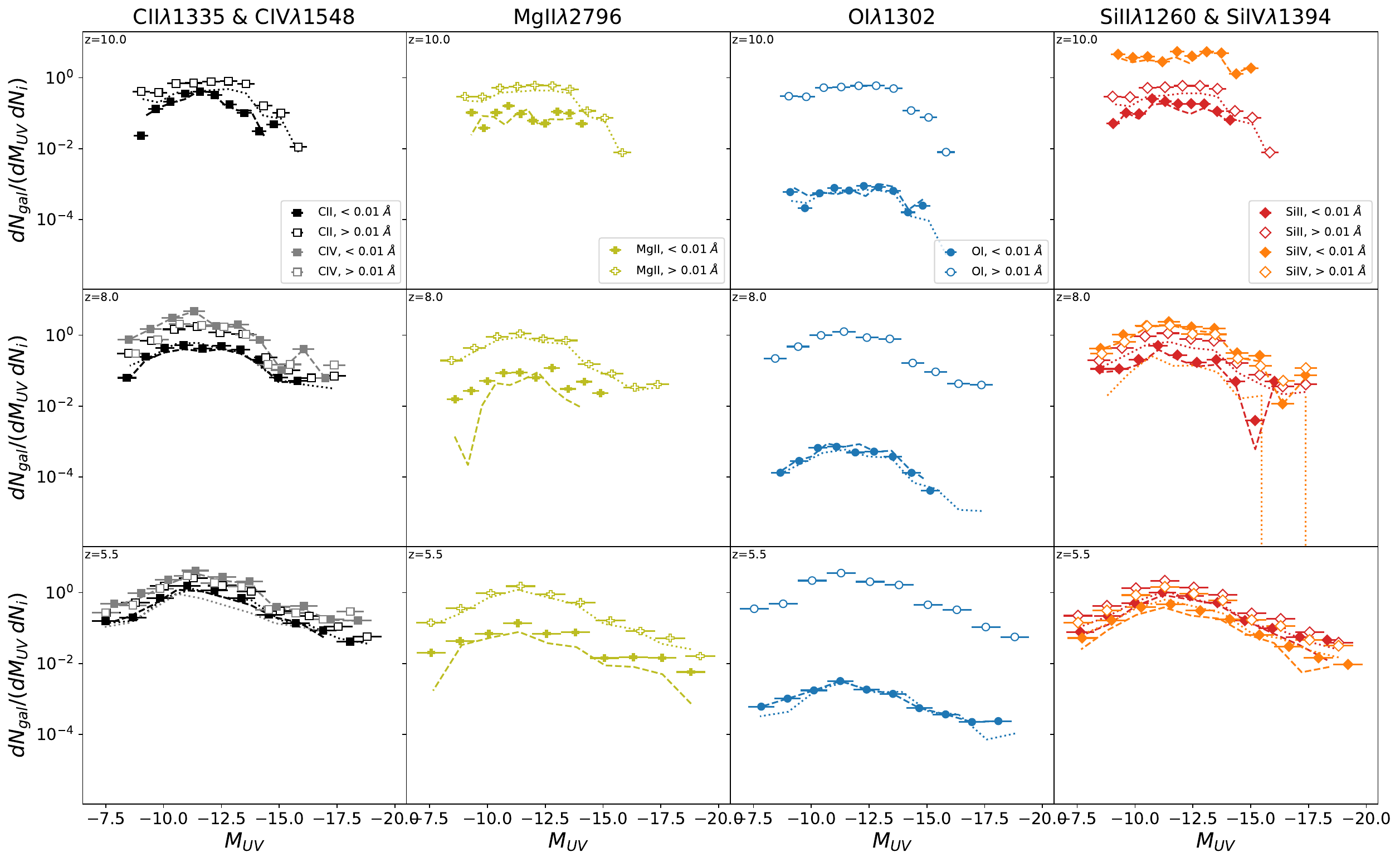}
    \includegraphics[width=1.0\textwidth]{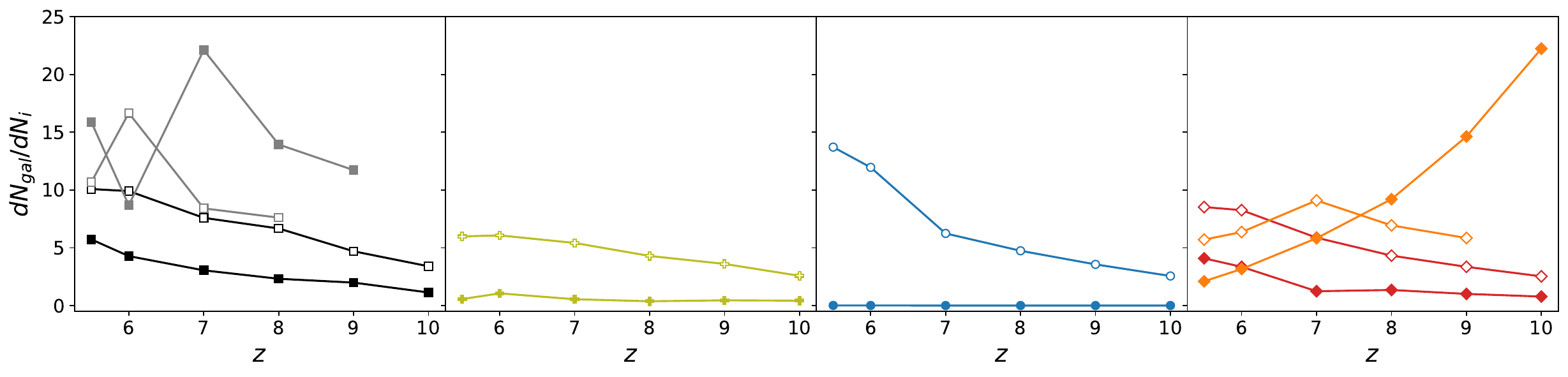}
    \caption{\textit{(top)} HGLD for each of the species throughout redshift, designated by the line for each species. The HGLD is the number of galaxies per magnitude bin per number of absorbers of a particular species. Each species is then split by an equivalent width of 0.01 $\mathrm{\AA}$. The dashed and dotted lines are further divided based on a quartile split on strength: 25\% and below (weaker) is dashed and 75\% and above (stronger) is dotted}. (bottom) Average number of galaxies per absorber vs. redshift, integrated from the \textit{top} HGLD and also split by 50\% EW. The UVLF for the higher equivalent width line for a species is larger overall, which means more galaxies are necessary to create the stronger absorber.
    \label{fig:UVLF_each_species}
\end{figure*}

While Figure \ref{fig:UVLF_all_W_pl12_CDF} indicates that, overall, the galaxies that contribute gas to metal absorbers are representative of the field population, this conclusion may depend on absorber strength. In order to test this, we divided up the absorber sample in Figure \ref{fig:UVLF_each_species}. This shows the species-specific UV HGLDs split at an equivalent width of $0.01 \mathrm{\AA}$, but still normalized by the total absorber count of a species.  We choose to split absorbers at an absorption equivalent width $EW = 0.01$ \AA~in order to split our simulated sample roughly in half. Observed absorbers tend to be stronger: compare, for example, the EW lower limit of 0.03\AA~of the XQR-30 catalog \cite{Davies:2023a}.

Figure~\ref{fig:UVLF_each_species} indicates that weak SiIV and CIV absorbers occur in rich environments at early times and poorer ones later on. By contrast, strong SiIV and CIV absorbers' environments do not evolve. The weak absorbers' evolution echoes the behavior seen in Figure~\ref{fig:CII_CIV_maps}, where CIV anti-correlates with galaxies at early times and is more closely associated with them later on.

Low-ionization absorbers uniformly trace richer environments as reionization proceeds because higher local gas densities are required in order to tilt the ionization balance towards lower ionization states in a reionized medium.
 
%\kfc{SAMIR The foregoing is an attempt at clarifying the discussion in the next (now commented-out) paragraph. Is that OK?}

%are unusual in that for a time period their stronger absorbers are more prominent among their absorber populations as seen in the bottom of Figure \ref{fig:UVLF_each_species}. High-ionization systems with weaker absorbers suggest richer environments as reionization progresses. The prevalence of stronger absorbers for high-ionization systems in these results suggest less overdense environments earlier on in cosmic time. If strong low-ionization absorber systems and weak high-ionization absorber systems mean a richer environment, then C and Si systems are around less overdense environments turning rich.

\begin{deluxetable}{lll}
    \tablenum{1}
    \tablecaption{KS test values between the HGLD and field UVLF at $z=10$.\label{tab:KS_all_species_10}}
    \tablewidth{30pt}
    \tablehead{
        \colhead{species} & \colhead{$K$} & \colhead{$p$}
    }
    \startdata
    CII & 0.2 & 0.994 \\
    CIV & 1.0 & 0.0   \\
    MgII & 0.2 & 0.994 \\
    OI & 0.2 & 0.994 \\
    SiII & 0.2 & 0.994 \\
    SiIV & 0.3 & 0.787
    \enddata
\end{deluxetable}

\begin{deluxetable}{lll}
    \tablenum{2}
    \tablecaption{KS test values between the HGLD and field UVLF at $z=5.5$.\label{tab:KS_all_species_5.5}}
    \tablewidth{30pt}
    \tablehead{
        \colhead{species} & \colhead{$K$} & \colhead{$p$}
    }
    \startdata
    CII & 0.1 & 1.0 \\
    CIV & 0.3 & 0.787   \\
    MgII & 0.2 & 0.994 \\
    OI & 0.2 & 0.994 \\
    SiII & 0.1 & 1.0 \\
    SiIV & 0.2 & 0.994
    \enddata
\end{deluxetable}

Figure~\ref{fig:UVLF_each_species} reveals at most small differences between the statistics of weak and strong absorbers, split by about 50\% between the absorber sample (points) and quartiles of the sample (lines). We use a KS test to compare the distributions of the strong absorbers and the weak absorbers with the null hypothesis being that the two distributions will be from the same distribution, and found no significance for all the species together. In summary, our simulations predict that all ions derive their metals from galaxies in a way that is unbiased with respect to luminosity. From Figure \ref{fig:UVLF_each_species} we do not see the absorber threshold strength having any preference towards any absolute magnitude. This can be seen visually as well as in the KS test results (Tables \ref{tab:KS_each_str_10} and \ref{tab:KS_each_str_5.5}). We have also tried comparing the environments of the lower and upper quartiles in the EW distributions, finding that the KS tests remain inconclusive. We conclude that the identity of the galaxy that ejected an absorber's metals cannot be inferred from the galaxy's luminosity or the absorber's EW.

\begin{deluxetable}{lll}
    \tablenum{3}
    \tablecaption{KS test values between the HGLD of the stronger absorbers and weaker absorbers at $z=10$. \label{tab:KS_each_str_10}}
    \tablewidth{0.5\textwidth}
    \tablehead{
        \colhead{species} & \colhead{$K$} & \colhead{$p$}
    }
    \startdata
    CII & 0.3 & 0.787 \\
    CIV & 1.0 & 0.0   \\
    MgII & 0.4 & 0.418 \\
    OI & 0.3 & 0.787 \\
    SiII & 0.3 & 0.787 \\
    SiIV & 1.0 & 0.0
    \enddata
\end{deluxetable}

\begin{deluxetable}{lll}
    \tablenum{4}
    \tablecaption{KS test values between the HGLD of the stronger absorbers and weaker absorbers at $z=5.5$.\label{tab:KS_each_str_5.5}}
    \tablewidth{0pt}
    \tablehead{
        \colhead{species} & \colhead{$K$} & \colhead{$p$}
    }
    \startdata
    CII & 0.1 & 1.0 \\
    CIV & 0.2 & 0.994   \\
    MgII & 0.3 & 0.787 \\
    OI & 0.3 & 0.787 \\
    SiII & 0.2 & 0.994 \\
    SiIV & 0.1 & 1.0
    \enddata
\end{deluxetable}

\subsection{Absorber Strength and Host Stellar Mass}
\begin{figure*}
    \centering
    \includegraphics[width=1.0\textwidth]{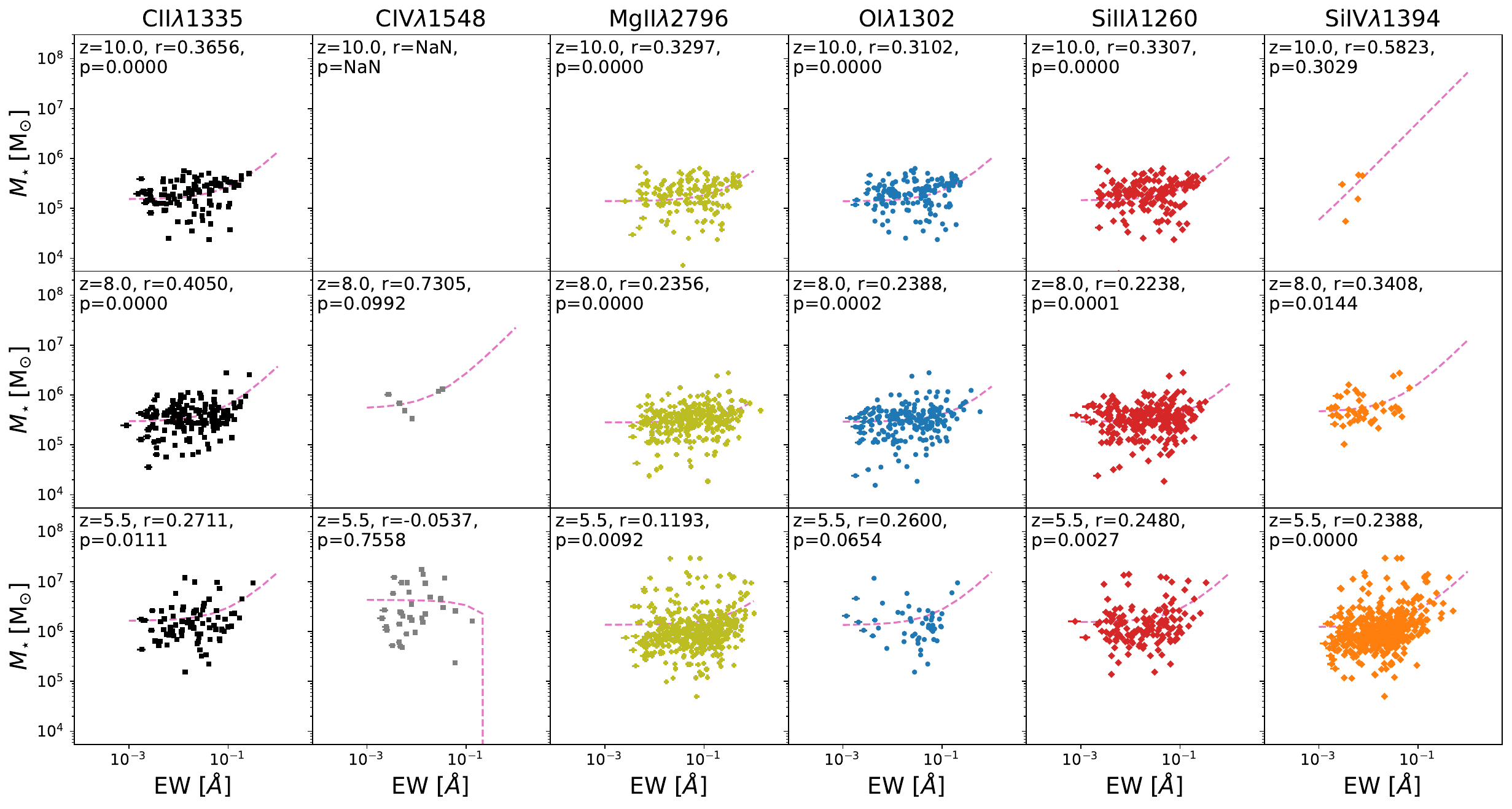}
    \caption{Scatter plots of the equivalent width of all absorbers related to mean stellar mass of all potential host galaxies for an absorber at redshift $z$ with line of best fit (pink, dashed) found from least-squared linear regression. The \textit{(r,p)} are Pearson coefficients and $p-$values respectively. Although weak, this suggests a positive correlation between EW and host stellar mass for most species aside for CIV, especially by $z=5.5$.}
    \label{fig:EW_Mstar}
\end{figure*}

Fundamentally, a galaxy's metal output reflects its stellar mass rather than its luminosity. For this reason, we now ask whether absorbers preferentially contain metals from massive or from low-mass galaxies. Figure \ref{fig:EW_Mstar} presents scatter plots of each absorber's equivalent width compared to the mean of its host galaxies' stellar masses, weighted by incidence as before. We also include least-square linear regression and Pearson coefficients. Despite significant scatter, we find that, for all species other than CIV, EW correlates positively with stellar mass. Hence absorbers' metals are predicted to be slightly more likely to originate in massive galaxies. 
%\kfc{These latter two sentences have the character of a ``caveat" that distracts and takes away from the result. Try moving them to a separate paragraph and, if possible, speculate as to what should be done about all that? For example, can you comment on whether these results change if you change the search box size or just use the nearest-galaxy approach?}

\begin{figure*}
    \centering
    \includegraphics[width=0.413\textwidth]{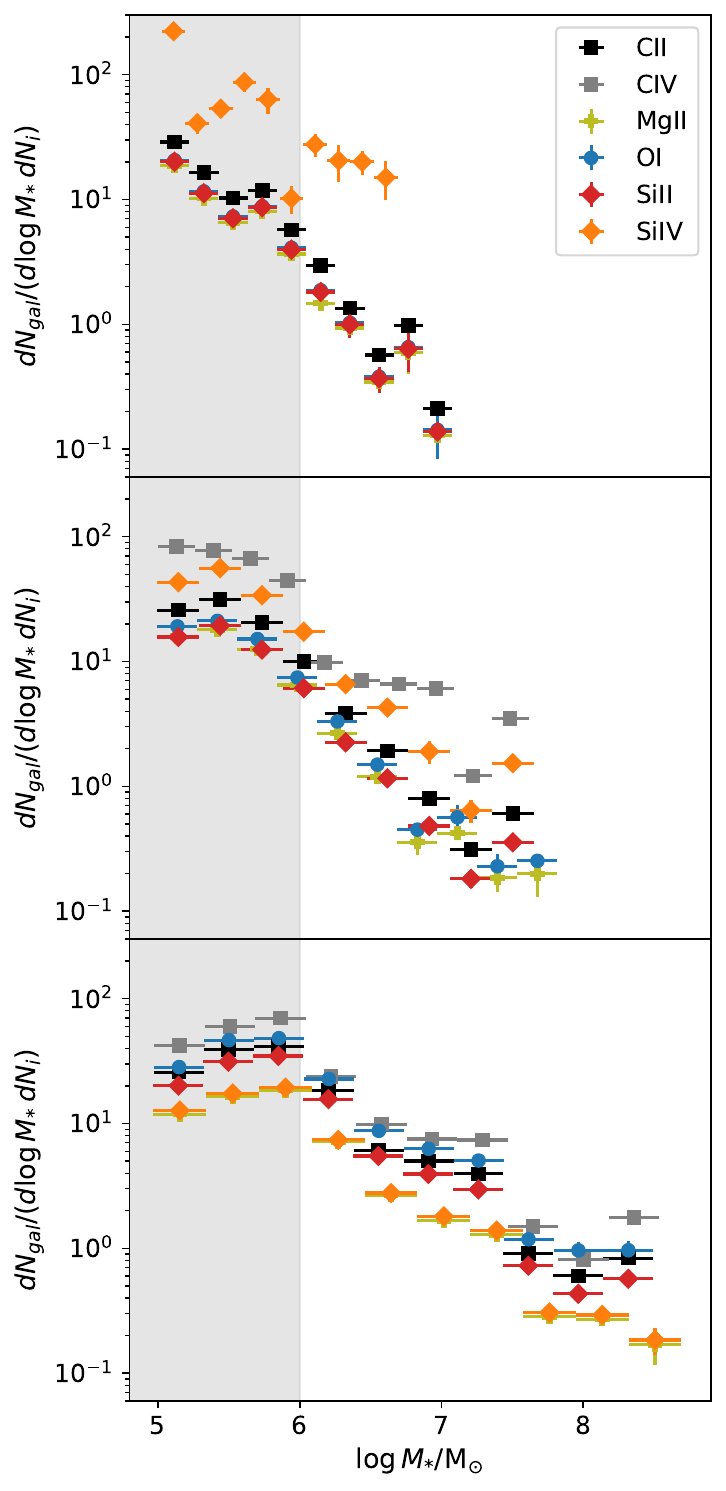}
    \includegraphics[width=0.4\textwidth]{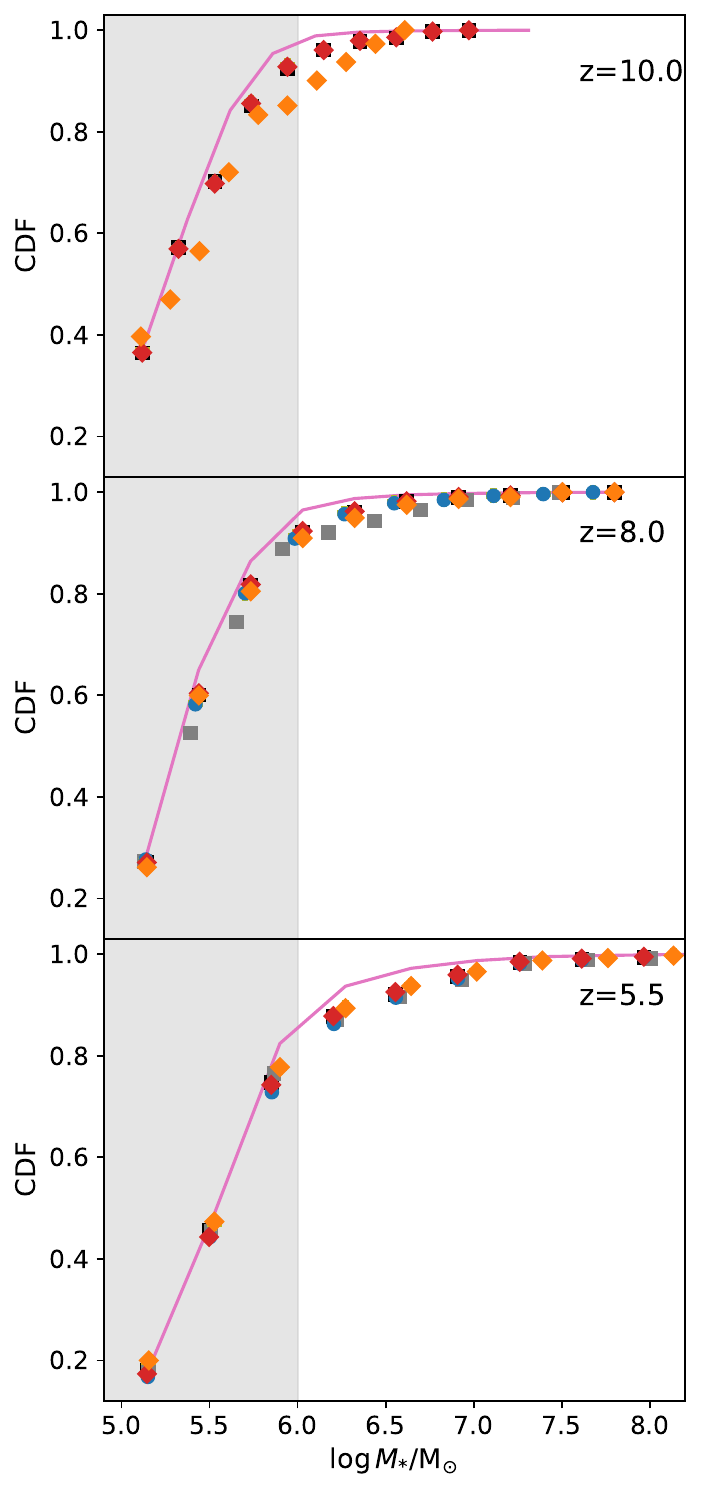}
    \caption{Host galaxy mass distributions (HGMDs) at three different redshifts (left). Comparison of cumulative HGMDs with the simulated field galaxy stellar mass function (SMF) (right). Similar to the HGLD, the HGMD is done with the same analysis with similar mathematical definitions as in Equation \ref{eqn:hgld_unnorm}, but by the galaxy's stellar mass instead of UV magnitude. The grey shaded region is the resolution limit of the simulation. A slight bias towards massive galaxies is apparent.}
    \label{fig:HGMD_all}
\end{figure*}

\subsection{Host Galaxy Mass Distribution} \label{sub:HGMD}
The weak positive correlations in Figure~\ref{fig:EW_Mstar} suggest that absorbers' host galaxy distributions show some dependence on stellar mass. In order to verify this, we reworked the the HGLD analysis to be done in bins of galaxy stellar mass rather than UV luminosity. Figure \ref{fig:HGMD_all} shows the resulting host galaxy mass distribution (HGMD) for various species in the same redshifts as in Figure \ref{fig:UVLF_all_W_pl12_CDF} as well as comparison to its field distribution, which is simply the (simulated) galaxy stellar mass function. This time, we do see a difference in the distributions, including in our statistically significant sample size absorbers such as CII, MgII, and SiII. In particular, the HGMD's CDF is boosted at the higher mass part, meaning higher fractions of the HGMD are higher mass galaxies. KS tests shown in Tables \ref{tab:KS_mass_all_10} and \ref{tab:KS_mass_all_5.5} for $z=10$ and 5.5 on the CDF in the right panel do not confirm a statistically significant difference between the HGMD and SMF shapes, but the KS test may not be a good metric for this analysis due to its limited range; we return to this in Section \ref{sec:discuss}. However, the p-values for these are lower and the test statistic is higher, suggesting that there is some difference in distribution, however insignificant.

\begin{deluxetable}{lll}
    \tablenum{5}
    \tablecaption{KS test values between the HGMD and field galaxy SMF at $z=10$.\label{tab:KS_mass_all_10}}
    \tablewidth{30pt}
    \tablehead{
        \colhead{species} & \colhead{$K$} & \colhead{$p$}
    }
    \startdata
    CII & 0.4 & 0.21 \\
    CIV & 1.0 & 0.0   \\
    MgII & 0.4 & 0.21 \\
    OI & 0.4 & 0.21 \\
    SiII & 0.4 & 0.21 \\
    SiIV & 0.5 & 0.084
    \enddata
\end{deluxetable}

\begin{deluxetable}{lll}
    \tablenum{6}
    \tablecaption{KS test values between the HGMD and field galaxy SMF at $z=5.5$.\label{tab:KS_mass_all_5.5}}
    \tablewidth{30pt}
    \tablehead{
        \colhead{species} & \colhead{$K$} & \colhead{$p$}
    }
    \startdata
    CII & 0.3 & 0.42 \\
    CIV & 0.3 & 0.42   \\
    MgII & 0.3 & 0.42 \\
    OI & 0.3 & 0.42 \\
    SiII & 0.3 & 0.42 \\
    SiIV & 0.3 & 0.42
    \enddata
\end{deluxetable}

\begin{figure*}
    \centering
    \includegraphics[width=\textwidth]{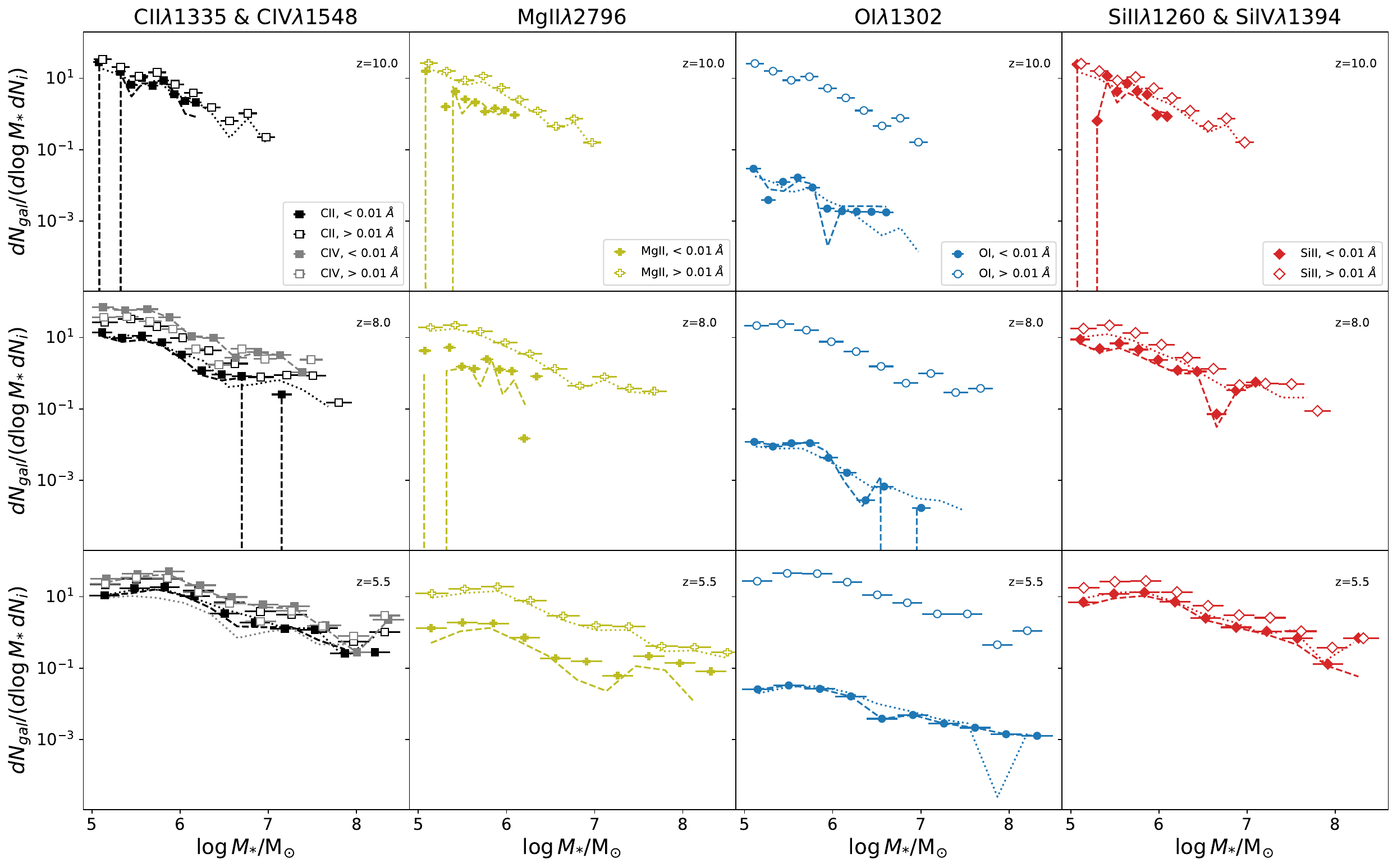}
    \caption{HGMD of each species split at different strengths. This has the same criteria as in Figure \ref{fig:UVLF_each_species}. The distributions do not appear to have any statistical difference, plus differences in CDF has high discrepency that averages close to 0 in all bins. This suggests not necessary most strong abosrbers are closer to more massive galaxies, just more metals come from more massive galaxies.}
    \label{fig:HGMD_each}
\end{figure*}

As in the case of the HGMD, we explored whether HGMD depends on absorber strength. Figure \ref{fig:HGMD_each} shows the HGMD of each species, subdivided by absorber strength as in Figure \ref{fig:UVLF_each_species}. These HGMD share similar shapes between each other, which the KS test could not significantly separate. In fact, the test statistic and p-values are similar between these HGMD and the HGLD of Table \ref{tab:KS_each_str_10} and \ref{tab:KS_each_str_5.5}. This suggests that stronger absorbers do not derive their metals from more massive galaxies even though there is a slight bias for more metals to originate in more-massive galaxies. However, more of the stronger absorbers' HGMD have bins that extend further into more massive galaxies rather than the weaker absorbers. Since the KS test only tests the range of values shared between distributions, it would only test on the mass range of the weaker absorbers' distributions, excluding higher mass data-points and making it incompatible for the original purposes. This should be reinterpreted as both less massive and more massive galaxies make the same amount of weak absorbers, but more massive galaxies are more capable of making stronger absorbers.

%\begin{itemize}
%    \item UV luminosity distribution presented: looking at each species and at particular strength $W = 0.01 \mathrm{\AA}$.
%    \item Do not see deviation between volume-based UVLF and species-based UVLF by much. Early on more so likely from smaller sampling. There is a slight preference towards lower mass galaxies for Si, , fontsize=12, fontsize=12
%    \item No differing magnitude (mass) between galaxies and absorbers: all low-ionization species folllow close to general trend
%    \item Weird things happening with CIV and SiIV. Likely now due to small sampling.
%    \item Boost at -12 mag for CIV and SiIV, stronger with CIV
%    \item SiIV flips expected galaxy LF in integrating to less than SiII at $z=5$; More SiIV can be made from galactic population than SiII. 
%    \item Begin to see deviation from species at -18 mag, so need more massive galaxies to see any particular magnitude favorability with species.
%    \item Do see for stronger absorbers, more galaxies needed to create it
%    \item except for CIV and SiIV from $8 > z > 6$, where more stronger absorbers are present
    
%\end{itemize}

\section{Discussion} \label{sec:discuss}
\iffalse
main takeaway: no dependence of metal contribution on luminosity
discussion 1: can't guess host based on absorber
discussion 2: consistency w/ previous results indicating most absorber hosts are low-mass
discussion 3: interaction with environment analyses
3a: rare/stronger absorbers imply richer environments
\fi
\subsection{Galactic population with Absorbers} \label{sub:gal_pop}
Our simulations predict that the metals in metal-absorbing gas originate from galaxies in a way that is unbiased with respect to galaxy luminosity; put differently, all galaxies contribute similar numbers of metal atoms to the observable absorber population. Given that faint galaxies are more abundant, they are therefore predicted to contribute most of the metals that are observed in absorption. This luminosity-independence undermines observational efforts to assign unique absorber-host identifications with absorbers in protocluster environments.

%\kfc{The draft makes the point in this last sentence several times. Are there observational papers that take this approach/make this assumption that can be referred to for motivation? Or is this already in the intro?} \skc{It's in the intro}

Switching from the observational to physical, a weak positive correlation in absorber strength versus mean host galaxy stellar mass suggests that absorbers' metals originate preferentially in galaxies with higher stellar mass. This discrepancy between HGMD and HGLD suggests the signal is weak, which includes the EW-stellar mass relation, or due to large scatter in the UV mass-luminosity relation.

The relatively weak relationships between stellar mass, absorber strength, and hence contributed metal mass are not strong enough to overcome the steep slope of the stellar mass function. Hence less-massive galaxies remain responsible for contributing most of the observable metals~\citep[][]{Finlator:2013,Finlator:2020, Doughty:2019}{}{}.

Consequently, it is more probable that any given absorber detection will trace the less massive galaxies more often than more massive galaxies. However, with no preferences between the species then one cannot infer any other properties of the galaxy from photometry. The lack of inference does not allow indirect relations and statistics on the galactic population.
Other results designate that SiIV and CIV tend to populate around higher Ly$\alpha$ luminosity and higher stellar mass overdensities before reionization ends \citep{Doughty:2023}. 
This is consistent in seeing our correlation of EW and host stellar mass for them before the end of reionization, but this does not include analysis discerning rich environents.
This also follows the slight correlation of EW and host stellar mass for these species before the end of reionization, yet these analyses might be impacted by rich environments that have many galaxies around absorbers.

%As such, how can we have all luminosities of galaxies contribute equally to absorbers yet have stronger absorbers near more massive galaxies? This appears to conflict with previous observations that equivalent widths at fixed distances are weaker for less massive halos \citep[][]{Churchill:2013}{}{}. One important difference to note is the discrepancy on getting mass from luminosity from given models. The noise in the mass-luminosity relation means that there is no immediate observable to test whether more massive galaxies are a strong absorber, or if strong absorbers are around a bright galaxy. As such, more work needs to go into observational analyses to correctly estimate the mass of a galaxy to see this; perform this with more than one photometric band at least. If one disregards the noise, two other hypotheses emerge that are not mutually exclusive.\kfc{I don't think I agree with this argument: noise tends to eliminate trends (for example, by creating massive galaxies in low-mass halos) rather than to create them artificially.}.

How can absorber strength correlate with host stellar mass, and therefore trace massive galaxies, but does not correlate with luminosity? Galaxies' halo masses impact absorber strength when observed at fixed distances, as found in \citet[][]{Churchill:2013}{}{} which have used photometry to infer halo mass. However, the key differences are the particular bands used, low-redshift relations potentially breaking down at high redshift, and larger noise between the stellar mass and luminosity relation in the UV. Otherwise, two physical hypotheses emerge that are not mutually exclusive.

One is these environments are definitely richer and are more probable in creating more massive galaxies with less massive galaxies as bound members of a proto-cluster or proto-group. As these galaxies are near each other, the gas ejected can mingle near the massive galaxies, potentially being gravitationally streamed into their CGM. These would allow higher density CGM around massive galaxies that are observed and still account for less massive galaxies' contributions. Such species are more prevalent in higher overdensities of $\Delta > 10$ \citep[][]{Turner:2016}{}{}. This hypothesis agrees with zoomed-in simulations around groups, where they find 45\% of gas within the massive galaxy is from its satellite halos \citep[][]{Grand:2019}{}{}.

Another hypothesis is that the cross section for massive galaxies are boosted for larger column densities when these galaxies are in higher density regions. The less massive galaxies would still contribute gas, but the more massive galaxies have their gas extended further to potentially form absorbers. Absorbers have been seen farther out from massive galaxies \citep[][]{Keating:2016}{}{}, but no modelling of the cross section to environment density has been done to our knowledge.

\subsection{Caveats} \label{sub:caveats}

TD resolves a diffuse UVB on a coarse grid. This raises questions on how this can impact our results. When calculating the absorbers, the radiative transfer grids are $80^3$ cells, or a side length corresponding to 150 comoving $h^{-1}$ kpc. This matches almost half the search radius for absorbers to galaxies, meaning that the grid is roughly one-quarter the space of the search volume. The regions are therefore sufficiently resolved for two halves of each CGM. With the linking radius being almost twice the grid length, this should be sufficient spatial resolution to be able to determine the general phase within the CGM and have the correct species for an absorber.

With this novel analysis, some concerns may arise for the weighing scheme. Primarily, are these good weights? First, using the number of incidences may not seem intuitive and it may be simpler to weigh it based on the closest galaxies. Although, this disregards the furthest distances that can be kinematically possible that was seen with Figure \ref{fig:gal_disp} as it practically assumes that the closest galaxy, even in a cluster, is the host. Additionally, the number of incidences also includes information on how many gas particles are associated with a galaxy, which would increase weights to more actively ejecting galaxies. Therefore, these weights reduce proximity bias of \textsc{hostgal} and consider more ejection-active galaxies as more probable.

We have tested alterations to \textsc{hostgal} to see if this had any impact on the analysis. First, we linearly displaced galaxies back given their velocities using the difference in time between the snapshot and a recorded last launch time for each gas particle and its corresponding LLP. However, the distance distribution of the nearest galaxy and LLP broadened compared to not reversing the galaxy positions, but the reverse assumed no immediate accelerations and simplifying the kinematics. Additionally, we also tested with just the nearest galaxy instead of potential hosts in a comoving box. The HGLD was very similar in shape and amplitude with the nearest galaxies compared to the HGLD of the potential hosts in a comoving box when looking at the species as seen in Appendix \ref{app:search_box}. The comoving box was preferred as the weighing scheme allows all kinematically possible hosts and points out more active galaxies.

There can be concern when looking at the numerical methods for the simulation and group finding. \textsc{SKID} and the synthetic photometry only require one star particle to be present in the snapshots to create the galaxy catalog and photometry respectively. With $512^3$ particle resolution and $12$ comoving $h^{-1}$ Mpc box length the simulation is capable of creating the lower-mass galaxies that are expected to be prevalent with some numerical deviations, such as the bend at the lower magnitudes of the HGLD. This is also why the HGLD are compared to the field UVLF calculated from the simulations rather than from analytical models. Increased dynamic range would significantly help with more fidelity creating more massive galaxies, but the current results show the first results of such an analysis with a significant number of galaxies needed for confidence.

The KS test is a capable test to see if there is a difference in distributions, but that is its limitations. The primary issue is that it cannot be used in a discrete population distribution \citep[][]{FrankMassey:1951}{}{}. We are limited in that we can only derive the distributions from our discrete sample. Additionally noticed from our analysis the null hypothesis that the distributions are identical needs more complete distributions. In our case, we had only parts of a sloped/Schechter distribution. Additionally, the test can only be done in shared ranges, which explains the similar values for the HGMD of different absorber strengths: it only tested lower mass bins. Finally, variations in the p-value between the HGLD and HGMD are at $\Delta p \sim 0.2$ with the average changes to the test statistic at $\Delta K \sim 0.1$, which is way too high to be able to tell if the p-value is significant. This suggests our current analysis and sample for the distributions are too low resolution for the KS test. This does not necessarily neglect the difference seen in the CDF, but no other statistical test can be performed at this moment due to the main analyzer's lack of statistical knowledge. Still, we suggest avenues to perform this in different ways, like the area between the CDFs \citep[similar to area under curve statistics,][]{BRADLEY19971145}{}{}. 
%\kfc{Can you give a reference for that approach?}
%\skc{No it's just an idea, not referencing anything as far as I know}

\section{Conclusion} \label{sec:summary}

%PUT CONCLUSION HERE. \lipsum[1-3]

%\ehc{I think I am biased, but I've always liked to see a clear bullet pointed list of conclusions, even if they're null results.}
%In conclusion, we do not see a preference for absorbers to galaxy UV magnitude nor stellar mass. We do not see certain galaxies with a preference to a species nor with preference to absorber strength. This challenges the paradigm of associating absorbers with the nearest observable galaxy within the field. 

We have performed an analysis linking the absorbers and galaxies within our simulation that is robust since it includes a vast majority of the potential galactic displacement, and tested to see if there is a correlation between species, absorber strength, and preferred galaxy mass or luminosity. We find that:

\begin{itemize}
    \item KS tests and by-eye comparisons between the field luminosity function and the HGLDs indicate that absorbers do not draw their gas preferentially from a particular galaxy luminosity range. This result is true both overall, and for absorber subsamples split at $0.01 ~\mathrm{\AA}$ (50\% split) and the weaker and stronger quartiles.
    \item A very weak correlation is found between absorber strength and weighted galaxy stellar mass. Although this exists in Figure \ref{fig:EW_Mstar}, we cannot see it directly in the HGLD of varying strengths in Figure \ref{fig:UVLF_each_species}, suggesting the weak signal.
    \item HGMD do see a difference from the galaxy SMF, with boosted fractions in higher mass galaxies. This suggests more metals are ejected by more massive galaxies.
    \item Although weaker absorbers are equally associated with less massive and more massive galaxies, more massive galaxies may be only capable of creating stronger absorbers.
    \item The large uncertainty in mass-luminiosity relations for galaxies do not allow immediate tracking of the UV luminosity of galaxies to their surrounding environment, washing out the relation if done in immediate photometry. A careful follow-up on finding galaxy stellar mass is more important.
\end{itemize}
%\begin{itemize}
%    \item There seems to be no preference to galaxy mass nor UV magnitude %No preference to certain absorber species and galaxy luminosity
%    \item No deviation from shape of UVLF of field
%    \item No deviation between strengths of absorbers except for abundance.
%    \item No relation between mass and EW. 
%\end{itemize}

%% Also note that the akcnowlodgment environment does not support long amounts of text. If you have a lot of people and institutions to acknowledge, do not use this command. Instead, create a new \section{Acknowledgments}.
\begin{acknowledgments}
%ACKNOWLEDGE HERE. \lipsum[1-1]
Samir Ku\v{s}mi\'{c} is supported by the National Science Foundation (NSF) under Award Number 2006550. The Technicolor Dawn simulations were enabled by the Extreme Science and Engineering Discovery Environment (XSEDE), which is supported by NSF grant number ACI-1548562, now transferred services to Advanced Cyberinfrastructure Coordination Ecosystem: Services \& Support (ACCESS). The Cosmic Dawn Center is funded by the Danish National Research Foundation. The analysis used \textsc{numpy}\citep[][]{Harris:2020}{}{} , \textsc{scipy}\citep[][]{Virtanen:2020}{}{} and \textsc{yt}\citep[][]{Turk:2011}{}{}.
\end{acknowledgments}


\vspace{5mm}
%\facilities{HST(STIS), Swift(XRT and UVOT), AAVSO, CTIO:1.3m, CTIO:1.5m,CXO}


%\software{astropy \citep{2013A&A...558A..33A,2018AJ....156..123A}, Cloudy \citep{2013RMxAA..49..137F}, Source Extractor \citep{1996A&AS..117..393B}}



\bibliography{references}{}
\bibliographystyle{aasjournal}



\appendix 

\section{Search Box Method vs. Nearest Galaxy} \label{app:search_box}

\begin{figure}[b]
    \centering
    \includegraphics[width=0.325\textwidth]{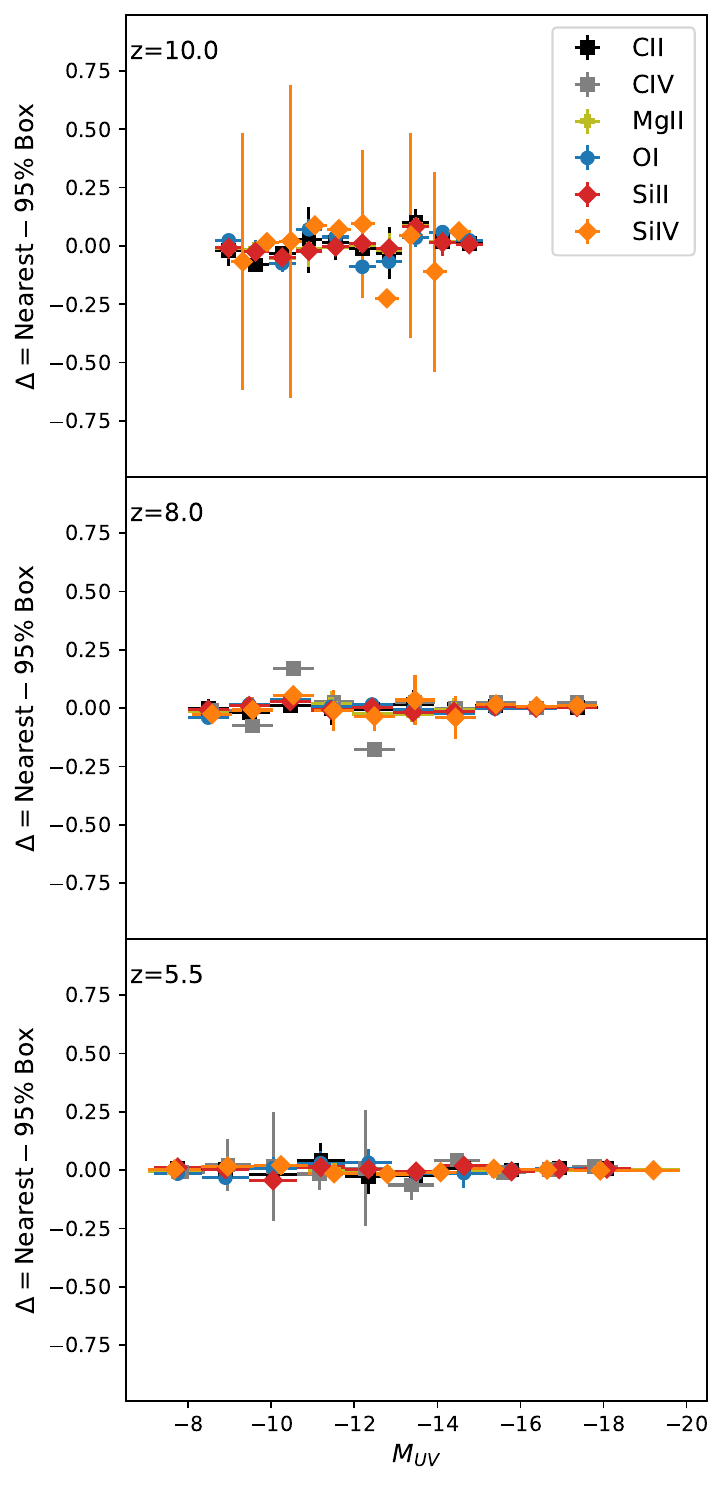}
    \includegraphics[width=0.325\textwidth]{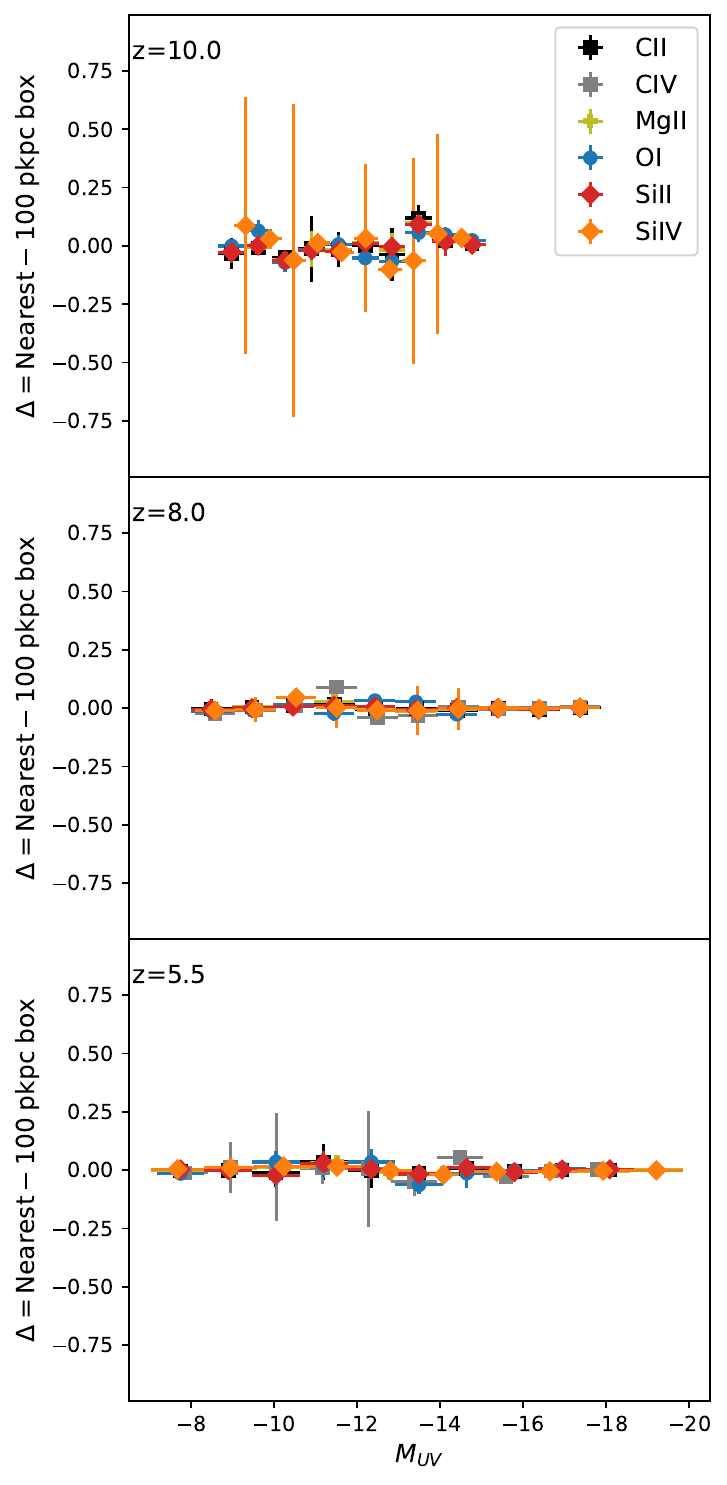}
    \includegraphics[width=0.325\textwidth]{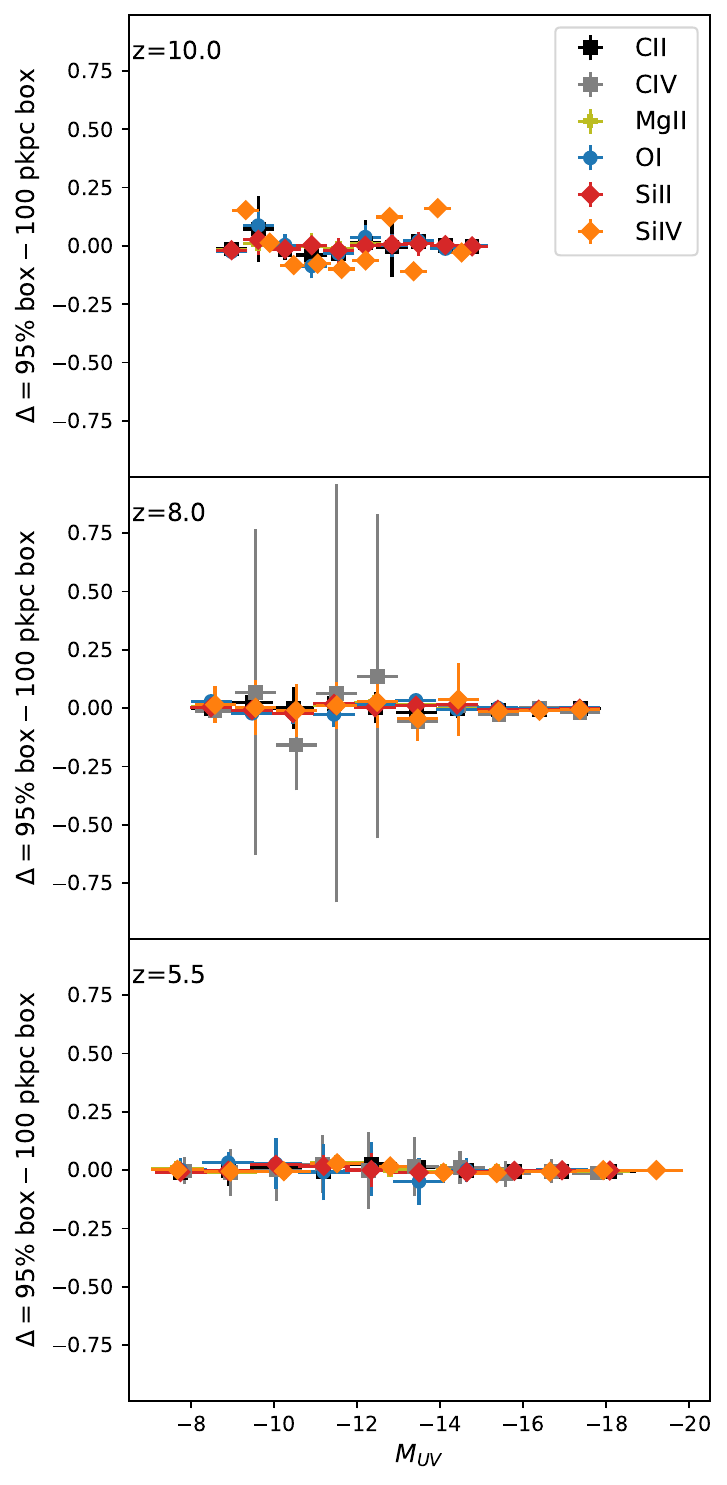}
    \caption{Difference between the normalized HGLD for each species between: the nearest galaxy approach vs. our 95\% box (left), nearest galaxy vs. 100 pkpc box (middle), and the 95\% box vs. 100 pkpc box (right). Errors are from the nearest galaxy method for (left) and (middle), 95\% box for (right). We do not see significant differences between the HGLD.}
    \label{fig:boxVsNearest}
\end{figure}

Figure \ref{fig:boxVsNearest} shows the difference between the normalized HGLD between our search box method and using the nearest galaxy. As can be seen, most points lay close to no difference with errors included. From this we conclude that no further systematic error comes from our search box method.

\end{document}